\documentclass{article}
\usepackage{spconf,amsmath,graphicx}
\usepackage{multirow}

\title{Streaming Bilingual End-to-End ASR model using Attention over multiple softmax}
\name{Aditya Patil \textsuperscript{1} \thanks{\textsuperscript{1}The work was done when Aditya was working at Microsoft}, Vikas Joshi, Purvi Agrawal, Rupesh Mehta}

\address{Microsoft Corporation, India \\
                        \ adityapatil$@$umass.edu \quad
                         \{vikas.joshi, agrawalpurvi, rupesh.mehta\}$@$microsoft.com}
\copyrightnotice{978-1-6654-7189-3/22/\$31.00~\copyright2023 IEEE}
\begin{document}
%
\maketitle
\begin{abstract}
Even with several advancements in multilingual modeling, it is challenging to recognize multiple languages using a single neural model, \textit{without knowing the input language} and most multilingual models assume the availability of the input language. In this work, we propose a novel bilingual end-to-end (E2E) modeling approach, where a single neural model can recognize both languages and also support switching between the languages, without any language input from the user. The proposed model has shared encoder and prediction networks, with language-specific joint networks that are combined via a self-attention mechanism. As the language-specific posteriors are combined, it produces a single posterior probability over all the output symbols, enabling a single beam search decoding and also allowing dynamic switching between the languages. The proposed approach outperforms the conventional bilingual baseline with 13.3\%, 8.23\% and 1.3\% word error rate relative reduction on Hindi, English and code-mixed test sets, respectively.

\end{abstract}
\section{Introduction}
\label{sec:intro}

Conversations among humans often involve using more than one language (multilingual) and mixing one language with another (code-mixing). The phenomenon is more predominant in multilingual regions like southern Asia where code-mixing and code-switching is observed in many scenarios including call center conversations, voice assistants and others. Also, with increasing adoption of voice assistants, users now expect a truly multilingual experience, where they can interact with voice assistants in any language, interchangeably, without explicitly setting the language of the conversation. Therefore, addressing the problem of multilingual and code-mixed speech recognition is increasingly becoming important. 

With the increasing demand of voice assistants in gadgets, the on-device ASR models are becoming popular as they alleviate the privacy concerns, can work without internet connectivity, and have low latency. However, providing a truly multilingual experience or handling code-mixing is especially challenging for on-device models. In the conventional hybrid systems, multilingual experience is provided by having multiple monolingual models in tandem and choosing the output of one of these models based on the decision of the language identifier (LID) \cite{joshi2021multiple}. Such an approach is not feasible for on-device models as the model size increases with the number of languages, while most of the on-device applications demand ASR models to have small memory footprint often restricted to $50-100$Mb. Moreover, such an approach is unable to handle code-mixed utterances effectively. Another approach to do multi-lingual modeling is by simply  pooling data and symbols from multiple languages and training the model \cite{tong2017investigation}. However, such a model has inferior performance compared to their monolingual counterparts, as reported in \cite{joshi2021multiple, Multilingual_RNNT}.

A prior effort in this direction proposed a MultiSoftmax model to improve the on-device recurrent neural network transducer (RNN-T) model performance, by having language-specific softmax, joint
and embedding layers, while sharing rest of the parameters \cite{joshi2021multiple}. The authors also extended the MultiSoftmax model to work without knowing the input language, by integrating a language identification
(LID) model, that estimates the LID on-the-fly and also does the recognition at the same time. Such class of multilingual models that do not require language input apriori are referred to as LID-free multilingual models. However, the LID-free MultiSoftmax model have the following two drawbacks:
\begin{itemize}
    \item Needs multiple beam search decoding (one per language) to be run in tandem, making it computationally expensive.
    \item Does not support code-mixed speech recognition as separate beam search is run for each language and there is no mechanism to switch from one language to another. 
\end{itemize}

In this work, we propose a novel  on-device bilingual model, with a single beam search that can recognize both languages and also support switching/mixing between the languages. This is achieved without any language input from the user and hence is a LID-free model. We propose multiple softmax with attention model with the following key components:
\begin{itemize}
    \item Shared encoder and prediction network.
    \item Language-specific output layers with softmax to produce posterior probabilities over corresponding language-specific output symbols.
    \item Attention mechanism that estimates weight of each language from the input frames.
\end{itemize}
The estimated attention weights are multiplied with the language-specific output posteriors and further concatenated to obtain a single vector with posterior probabilities over the combined symbol set. Therefore, the proposed model is an extension to MultiSoftmax model previously proposed in \cite{joshi2021multiple} with the following benefits: 
\begin{itemize}
    \item Having a single output vector with  posterior probabilities over all language symbols, which enables a single beam search decoding, instead of multiple language-specific beam search decoding.
    \item It also allows mixing between the languages as beam search can select output symbols from any language at any given instant.
    \item Attention weights estimate the input language and guide the beam search decoding appropriately.
\end{itemize}

The proposed approach is applicable to any on-device ASR models such as recurrent neural network transducers (RNN-T) or transformer transducer (TT). We conduct experiments with RNN-T models and show $13.3\%$, $8.23\%$ and $1.3\%$ WERR reduction over conventional bilingual model on Hindi, English and code-mixed test sets. The analysis of the attention weights show the ability of the proposed approach to capture language information and in-turn improve the model performance.

The rest of the paper is organized as follows: We discuss prior work in section~\ref{sec:prior_work} followed by discussion of the proposed approach in Section~\ref{sec:proposed_approach}. We next discuss the experimental set-up and results in Section~\ref{sec:expt_details} and Section~\ref{sec:discussion_res}, respectively. The analysis of attention weights is discussed in Section~\ref{sec:analysis_weights} followed by conclusions in Section~\ref{sec:conclusion}.

\section{Relation to Prior Work}
\label{sec:prior_work}
Transfer learning, multilingual and multi-dialect modeling are widely used in hybrid ASR systems\cite{MLT_1,swietojanski2012unsupervised, MLT_2,TL_1,TL_2,Scanzio-MultisoftmaxFirstPaper,SHL_1,SHL_2,Seltzer-MTLPhonemeRecog,EL_1, EL_2,Das-MultiDialectEnsemble}. Transfer learning \cite{TL_1,TL_2} methods leverage a well trained acoustic model (AM) from high-resource language to bootstrap the target AM. A natural extension is to train multilingual model \cite{Scanzio-MultisoftmaxFirstPaper,SHL_1,SHL_2,Seltzer-MTLPhonemeRecog}, by using data from multiple languages to build robust seed model. 

Similarly, transfer learning \cite{Joshi2020, giollo2020bootstrap} and multilingual modeling is proposed for E2E ASR models as well in~\cite{MLT_RNNT_Umesh,joshi2021multiple, Multilingual_RNNT, Amazon_BilingualLID,pratap2020massively,Zhu2020MultilingualSR}. A streaming multilingual E2E model with language-specific adapters is proposed in~\cite{Multilingual_RNNT}. They show significant improvements by using language-specific one-hot vector. Similarly, Transformer-Transducer  based multilingual model is proposed in~\cite{Zhu2020MultilingualSR}. Several other multilingual models are proposed in the literature that which often better than monolingual models~\cite{bytes_E2E,pratap2020massively} or serve as robust seed models to further train the monolingual models. However, these methods are not LID-free, i.e., they need to know the input language apriori. Therefore, the focus of this work is to build LID-free multilingual models. 

LID-free multilingual models are recently proposed in~\cite{MLT_RNNT_Umesh} and~\cite{Amazon_BilingualLID}. Authors in~\cite{MLT_RNNT_Umesh} propose to use LID tokens at the start and end of the utterance, however, this model is proposed in the context of S2S models and is not streaming in nature.  Use of acoustic LID embeddings along with language tokens instead of using a predefined one-hot LID is proposed in~\cite{Amazon_BilingualLID}. Our method differs from the above method by not using LID embeddings but implicitly estimate the language-specific weight via an attention mechanism. Our work is closely related to the MultiSoftmax model proposed in~\cite{joshi2021multiple}. We discuss the difference between our proposed approach and the MultiSoftmax model in detail in the next section.

\begin{figure*}
\begin{center}
\includegraphics[trim={0.6cm 0 0 0}, clip, scale=0.62]{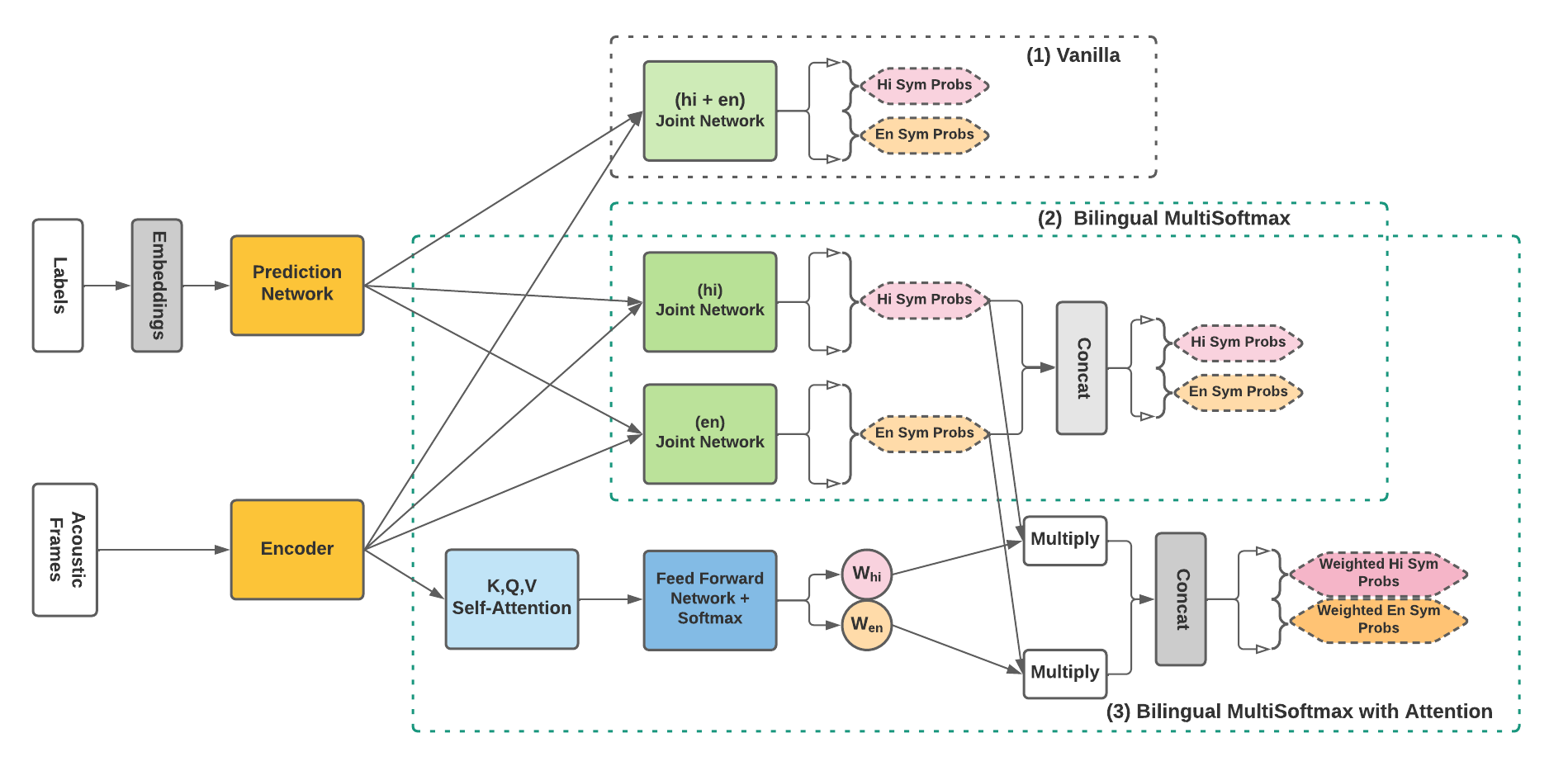}
\end{center}
\vspace{-0.4cm}
\caption{Block schematic of (1) Vanilla bilingual model, (2) Bilingual MultiSoftmax model, and (3) Bilingual MultiSoftmax with Attention model.}
\label{fig:blockDiag}
\end{figure*}

\section{Proposed Approach}
\label{sec:proposed_approach}
The RNN-T model consists of encoder network, prediction network and joint network~\cite{RNNT_graves}. The encoder transforms the input acoustic feature vector into latent representations. The prediction network transforms the previously predicted non-blank symbol into latent representations, suitable to predict the next symbol. The
joint network along with softmax nonlinearity combines the encoder and
prediction network representation to produce posterior probabilities over the symbols. In this work, we discuss \textit{three} different bilingual models and all of them have shared  encoder and prediction network, and only differ in the way joint networks are configured and also in the computation of posterior probabilities over output symbols. Fig.~\ref{fig:blockDiag} shows three different bilingual RNN-T model architectures namely: [1] Vanilla bilingual model, [2] Bilingual MultiSoftmax, and [3] Bilingual MultiSoftmax with Attention. We next discuss these three bilingual model architectures in detail.

\subsection{Vanilla bilingual model}
\label{sec:mRNN-T}

The vanilla bilingual model has all  parameters shared including the joint network, as shown in Fig.~\ref{fig:blockDiag}(1). The softmax is applied over a combined symbol set, which is a union (concatenation) of English and Hindi symbols. The  model is trained by simply pooling the data from both languages. Inference is done using single beam search  on the combined symbol set. The vanilla bilingual model is simple, and yet is a streaming  model which recognizes both languages without needing the language input from user. The vanilla model serves as a baseline for our proposed bilingual models.

\subsection {Bilingual MultiSoftmax model}
\label{sec:no_attn}
The bilingual MutliSoftmax model has language-specific joint  network and softmax layers, while shares the rest of model parameters, as shown in Fig. \ref{fig:blockDiag}(2). These joint networks along with softmax produces posterior probabilities over each language-specific symbol set. These posterior probabilities are concatenated to obtain a single vector consisting of posterior probabilities over entire symbol set.

The proposed bilingual MultiSoftmax model is similar to MultiSoftmax model proposed in \cite{joshi2021multiple} and differs only in terms of concatenating the posterior probabilities over two language-specific output symbols. Though the change is simplistic in nature, it is crucial as it now allows a single beam search over combined symbol set (instead of multiple beam search, one per language) and also enables code-mixing as the beam search can now dynamically switch between different language symbols.

\subsection{Bilingual MultiSoftmax with Attention}
\label{sec:mutli_softmax_attn}

We next introduce an attention mechanism over the bilingual MultiSoftmax model as shown in Fig. \ref{fig:blockDiag}(3) and is referred to as bilingual MultiSoftmaxAttn model. Similar to bilingual MultiSoftmax model, it has shared encoder and prediction networks along with language-specific joint networks followed by softmax nonlinearity. It also concatenates the language-specific output probabilities, albeit after multiplying with the corresponding \textit{language-specific} weights as shown in Fig.~\ref{fig:blockDiag}(3). The language-specific weights are estimated via an attention mechanism consisting of self-attention block along with feed-forward neural network and a softmax layer. Specifically, we estimate two weights, $w_{en}$ and $w_{hi}$ representing the weight to be multiplied to English (\textit{en}) and Hindi (\textit{hi}) posterior probabilities, respectively. The attention mechanism is analogous to language identification system and the weights, $w_{en}$ and $w_{hi}$, correspond to the probability of input language being English and Hindi, respectively. After multiplying the weights, the weighted posteriors are concatenated to obtain a single vector with posterior probabilities over the combined symbol set.

Even though we do not explicitly provide the language information, the model implicitly learns from the transcriptions as the words (and hence the labels) are represented with the corresponding language script. English words are transcribed in Latin script and Hindi words are transcribed in Devanagari script. Hence, an English utterance will have words transcribed in Latin only, a Hindi utterance will have words transcribed in Devanagari only and a code-mixed utterance will have mixed script with English words transcribed in Latin and Hindi words transcribed in Devanagari, respectively. Therefore, the model learns to produce English words in Latin script and Hindi words in Devanagari script, thereby preserving the original script.     

Apart from incorporating the language information via attention, another motivation to introduce the attention mechanism is to allow the network to leverage language information from contextual frames and incorporate it to estimate the posterior probabilities. For this purpose, the self-attention model uses all the frames in the past and few frames in the future, determined by the \textit{look-ahead} factor. For example, look-ahead of $10$ would imply using $10$ frames in future and all the frames in the past. Hence, if we are looking for only $10$ frames in future, it is still a streaming model similar to latency controlled BLSTM~\cite{LC_BLSTM} or transformer transducer with look-ahead. Infinite look-ahead would mean that all the future frames are used in attention weight estimation and model is not streaming in nature.

\noindent \textbf{Training and inference}: We train the model in the following three steps:
\begin{itemize}
    \item We first train a MultiSoftmax model~\cite{joshi2021multiple} with English and Hindi data.
    \item Next, we train the bilingual MultiSotmax model with English, Hindi and code-mixed utterances.
    \item Finally, we train bilingual MultiSotmaxAttn model with English, Hindi and code-mixed utterances
\end{itemize}
The inference is done via a single beam search decoding over combined symbol set, as done in case of the vanilla model.

\section{Experimental details}
\label{sec:expt_details}

\subsection{Data}
\label{data}
We conduct experiments on English (En-IN), Hindi (Hi-IN) and Hindi-English (Hinglish) code-mixed utterances. We use approximately $20.6K$ hours of training data consisting of $~8K$ hours of Hindi,  $~10K$ hours of English and $~2.6K$ hours of code-mixed data. The test set consists of $12240$ Hindi utterances, $25771$ English utterances and $3161$ code-mixed utterances. The train and test transcriptions are transcribed with the respective language scripts. Hence, English words are written in Latin and Hindi words are written in Devanagari. The output vocabulary consists of unique graphemes from the respective languages. Both languages have distinct set of graphemes. For every symbol, we also include $B$\_ prefix symbol in order to convert grapheme sequence to word sequence. Therefore, the final vocabulary consists of original graphemes, graphemes with  $B$\_ prefix, $<\text{blank}>$ symbol and  symbols for noise/fillers. In total, we have $59$ symbols for English and $133$ symbols for Hindi.


\subsection{Experimental setup}
\label{expt_setup}

We use $80$-dimensional log mel filterbank (LFB) features, computed every $10$ milliseconds (ms). These $80$ dimensional features are stacked to form $680$ dimensional features. The frames are sampled with a sampling factor of $3$ to obtain $680$ dimensional features for every $30ms$. The encoder is a $6-$layer unidirectional LSTM model with $1024$ hidden units. The prediction network is a $2-$layer unidirectional LSTM with $1024$ hidden units. Models are trained with Adam optimizer and with $32$ GPUs. Warm-up is done on $8$ GPUs for one sweep of the data before training with $32$ GPUs. All models are trained using the PyTorch toolkit~\cite{PyTorch}. The experiment set-up is analogous to other end-to-end research works~\cite{joshi2021multiple,Li2020Developing, Li2020Comparison, Li2019RNNT}.





\begin{table}
\caption{\textit{WER$[\%]$ comparison for Vanilla, bilingual MultiSoftmax and bilingual MultiSoftmaxAttn models on Hindi, English and code-mixed test sets.}}
\vspace{0.1cm}
\resizebox{8.6cm}{!}{
\begin{tabular}{|c|c|c|c|c|}
\hline 
 {\multirow{2}{*}{}} &  {\multirow{2}{*}{Vanilla}} & {{Bilingual}} & {{Bilingual}}\\
 & & MultiSoftmax & MultiSoftmaxAttn
 \tabularnewline
\hline 
\hline 
Hi-IN   & 17.37 & 14.81 & 15.05\tabularnewline
\hline 
En-IN   & 19.79 & 18.98 & 18.16 \tabularnewline
\hline 
Code-Mixed   & 31.80 & 33.24 & 31.37\tabularnewline 

\hline 
\end{tabular}
}
\label{tab:msa_wer}
\end{table}


\begin{table}
 \caption{\textit{WER$[\%]$ comparison for look-ahead $10$ and infinity (all the frames in the utterance) with bilingual MultiSoftmaxAttn model.}}
 \vspace{0.1cm}
 \centering
 \begin{tabular}{|c|c|c|c|c|c|}
\hline 
 &  Look-ahead 10 & Look-ahead inf. \tabularnewline
\hline 
\hline 
Hi-IN & 15.05 & 14.68  \tabularnewline
\hline 
En-IN & 18.16 & 16.56 \tabularnewline
\hline 
Code-Mixed & 31.37 & 31.69 \tabularnewline 
\hline 
\end{tabular}
\label{tab:la}
\end{table}
\begin{figure}[t]
	\begin{minipage}[b]{1.0\linewidth}
		\centering
		\centerline{\includegraphics[trim={0.5cm 0 0 0}, clip, width=8.52cm]{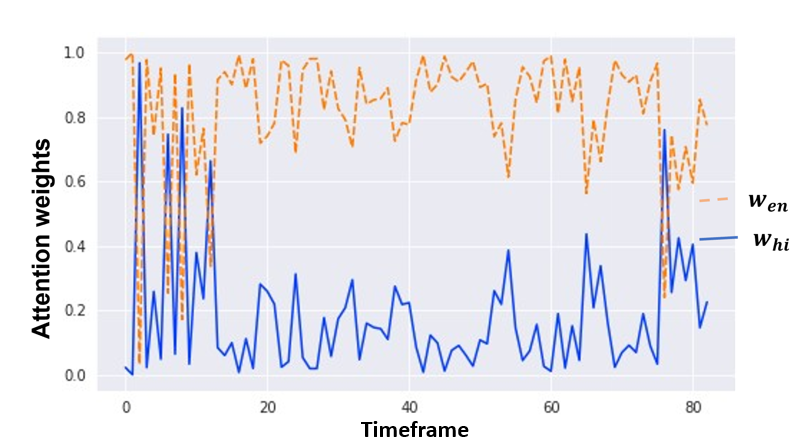}}
		\centerline{(a) English Utterance}\medskip
	\end{minipage}
	\begin{minipage}[b]{1.0\linewidth}
		\centering
		\centerline{\includegraphics[trim={0.1cm 0 0 0}, clip, width=8.85cm]{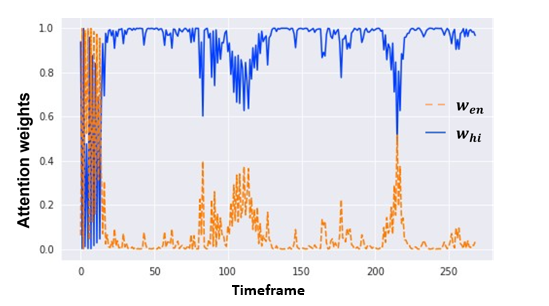}}
		\centerline{(b) Hindi Utterance}\medskip
	\end{minipage}
    \vspace{-0.5cm}
	\caption{Attention weights over time frames for (a) English, and (b) Hindi utterance, respectively.}
	 \label{fig:attnwts_enhi}
\end{figure}

\begin{figure}[t]
	\begin{minipage}[b]{1.0\linewidth}
		\centering
		\centerline{\includegraphics[trim={0.15cm 0 0 0.4cm}, clip, width=9.1cm]{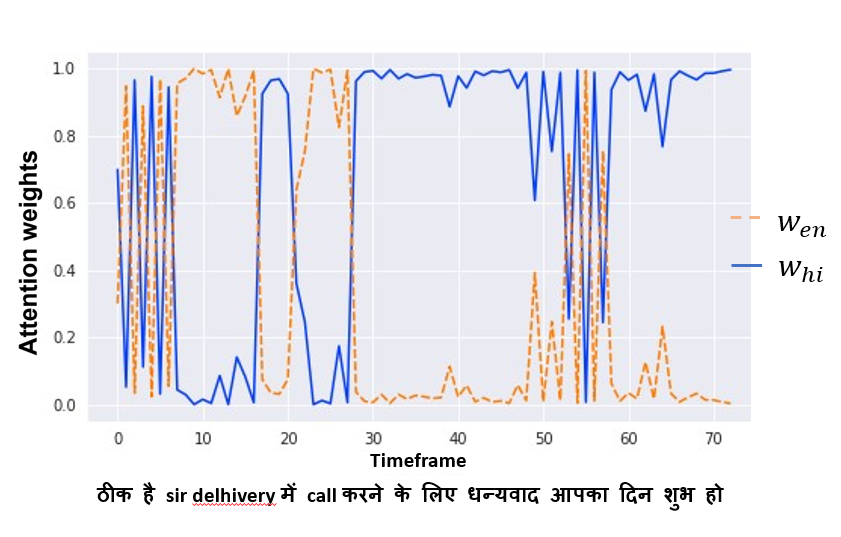}}
	\end{minipage}
	\begin{minipage}[b]{1.0\linewidth}
		\centering
		\centerline{\includegraphics[trim={0.2cm 0 0 0},clip, width=9.1cm]{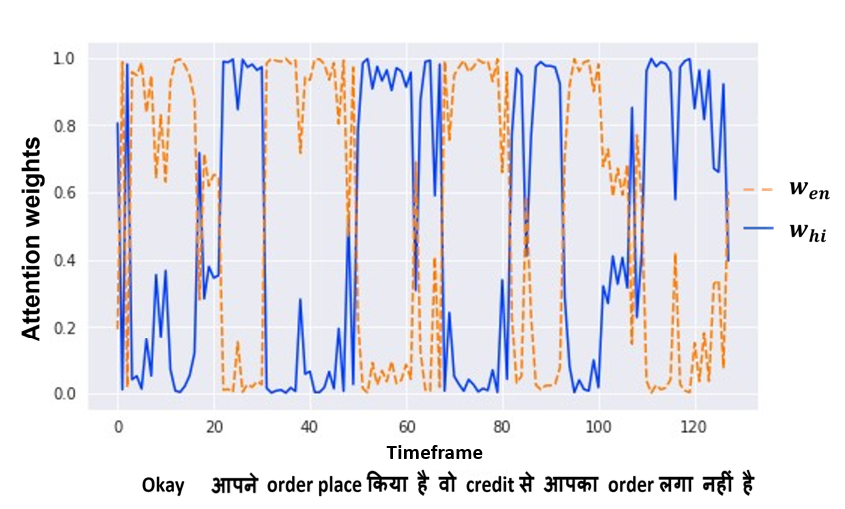}}
	\end{minipage}
	\caption{Attention weights over time frames for code-mixed utterances.}
	\label{fig:attnwts_CM}
\end{figure}
\section{Discussion of results}
\label{sec:discussion_res}
Table~\ref{tab:msa_wer} shows ASR system WER for Hindi, English and code-mixed test sets for vanilla bilingual model (baseline) and the proposed approaches, i.e., bilingual MultiSoftmax and bilingual MultiSoftmaxAttn. The bilingual MultiSoftmax performs better than vanilla model with $14.7\%$ and $4\%$ WERR reduction on Hi-IN and En-IN test sets, respectively. However, its shows $4.5\%$ WERR degradation on the code-mixed test sets compared to the vanilla model. With the introduction of attention, the bilingual MultiSoftmaxAttn model performs better than vanilla model with $13.3\%$, $8.23\%$ and $1.3\%$ WERR reduction on Hindi, English and code-mixed test sets. The results indicate that attention mechanism helps improve the model performance especially on code-mixed utterances.

The WER reported for bilingual MultiSoftmaxAttn in Table~\ref{tab:msa_wer} is with look-ahead of $10$ frames, i.e. the attention weights are estimated with $10$ frames in future and all the frames in the past. Note that in our experiments, $10$ frames in future are used only to estimate the attention weights. The additional look-ahead frames can be used to improve the encoder model performance by using latency-controlled BLTSM \cite{LC_BLSTM} or with transformer transducer with look-ahead, and we plan to experiment with that in future. Table~\ref{tab:la} shows the WER comparison for look-ahead of $10$ frames and all future frames in the utterance, referred to as look-ahead infinity. As shown in Table~\ref{tab:la}, WER improves on En-IN and Hi-IN with look-ahead infinity, while there is small regression on code-mixed test sets.
A probable reason for not observing the necessary gains could be due to fluctuations in the attention weights across adjacent frames. For an example utterance, $w_{en}$ values were observed to be $[0.8, 0.2, 0.9, 0.1]$ for frames $t$ to $t+3$, which implies that the attention model is changing its preference from English to Hindi for those adjacent frames. However, this is not possible in practice as each frame corresponds to $30ms$ only. Such fluctuations can adversely impact the beam-search decoding and we plan to address this in future by smoothing the attention weights.

\begin{figure}[!t]
\centering
\includegraphics[trim={0 0 0 0},clip, width=7.3cm]{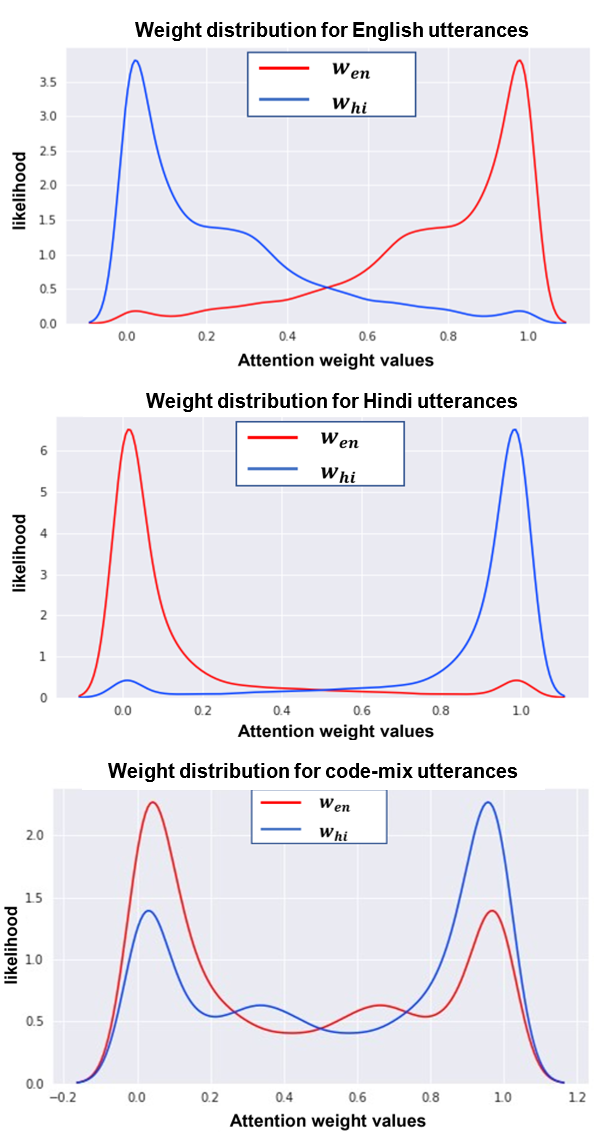}
\vspace{-0.1cm}
\caption{Plot of probability density function (pdf) for attention weights for English, Hindi and code-mixed utterances, respectively. The weight values on x-axis extend beyond and $0$ and $1.0$ as we plot the entire multi-modal Gaussian distribution estimated over the weight values. The y-axis represents the likelihood of the pdf.}
\label{fig:ms_attn_pdf}
\end{figure}


\section{Analysis of attention weights}
{\label{sec:analysis_weights}}
 
In this section, we analyse the attention weights ($w_{en}$ and $w_{hi}$) estimated for English, Hindi and code-mixed utterances. Fig.~\ref{fig:attnwts_enhi} shows the estimate of attention weights over time for a randomly sampled English and Hindi utterance. As can be seen, $w_{en}$ is close to $1.0$ for most of the time frames in English utterance and $w_{hi}$ close to $1.0$ for Hindi utterance. It is interesting to note that the attention weights fluctuate for initial frames indicating that the attention model is not confident in the initial frames. Such fluctuation can adversely affect beam search decoding and we plan to address this issue in future by employing smoothing over attention weights or by have custom logic to apply attention only after they are confident in their estimates.

Fig.~\ref{fig:attnwts_CM} shows attention weights over time for two sample code-mixed utterances. The corresponding transcription is also shown below the plot, with English words written in Latin and Hindi words written in Devanagari script. Comparing the trajectory of the attention weights and the corresponding transcriptions, it is interesting to note that $w_{en}$ is higher when English words are spoken and $w_{hi}$ is higher when Hindi words are spoken, thereby showing the efficacy of the attention based approach. 

Fig.~\ref{fig:ms_attn_pdf} shows the probability density function (pdf) for the attention weights estimated on English, Hindi and code-mixed utterances. For each set of utterances (English, Hindi and code-mixed), the attention weights are estimated over all the frames and we fit a multi-modal Gaussian distribution over the weights. As seen from Fig.~\ref{fig:ms_attn_pdf}, $w_{en}$ peaks near $1.0$ for English utterances and $w_{hi}$ peaks near $1.0$ for Hindi utterances. The  English distribution is more flat with small peak near $0.5$ and we hypothesize the reason to be the presence of  entities such as movie names, song names, person names and places in the English utterances, which cannot be  classified as English only.  In case of code-mixed utterances, $w_{en}$ and $w_{hi}$ peak at $0$ as well as $1.0$, indicating the presence of Hindi and English words. Hindi is the primary language in the code-mixed utterances and hence $w_{hi}$ has larger peak at $1.0$ compared to $w_{en}$. 

In summary, the analysis of attention weights show that it conforms to our initial hypothesis of attention model behaving like an implicit language identification system, where $w_{en}$ is close to $1.0$ for English words/frames and $w_{hi}$ is close to $1.0$ for Hindi words/frames. It also showed scope for improvement, especially in the initial frames.

\section{Conclusions}
\label{sec:conclusion}
The key conclusions from the paper are summarized as follows:
\begin{itemize}
    \item Proposed streaming, LID-free, on-device E2E ASR model, capable of recognizing of bilingual and code-mixed speech.
    \item Specifically, we proposed bilingual MultiSoftmaxAttn model, which combines language-specific symbol probabilities via a self-attention mechanism and produces a single posterior probabilities over the combined symbol set, by concatenating the weighted language-specific posteriors. The model leverages the idea of sharing most of the model parameters for robust parameter estimation, while have some language-specific parameters to learn language-specific characteristics.
    \item The language-specific components are obscured with the help of self-attention mechanism and concatenation operation which enables the model to produce a single  probability vector over all symbols and have a simplified beam search decoding as in case of the vanilla bilingual model.
    \item The proposed approach improves  over the vanilla bilingual baseline with $13.3\%$, $8.23\%$ and $1.3\%$  WERR reduction on English, Hindi and code-mixed test sets, respectively.
    \item Analysis of attention weights for monolingual and code-mixed utterances shows the efficacy of the proposed approach to capture language characteristics.
\end{itemize}
In future, we plan to work with transformer transducer model and explore smoothing of the attention weights to avoid random fluctuations in the estimated attention weights.

\bibliographystyle{IEEEbib}
\bibliography{strings,refs}

\end{document}